\documentclass[times, doublespace]{qjrms4}

\usepackage[colorlinks,linkcolor=blue,bookmarksopen,bookmarksnumbered,citecolor=blue,urlcolor=blue]{hyperref}
\graphicspath{{Figures/}}

\usepackage{moreverb}
\usepackage[switch]{lineno}
\usepackage{comment}
\usepackage{esint}
\usepackage{mathtools}
\usepackage{anyfontsize}

\newcommand{\U}[1]{\ensuremath{\mathrm{#1}}}
\newcommand{\sav}[1]{\left\langle {\smash{#1}} \right\rangle}
\newcommand{\spav}[1]{\left\langle {\smash{#1}} \right\rangle_{xy}}
\renewcommand{\vec}[1]{\boldsymbol{#1}}
\renewcommand{\d}{\textnormal{d}}
\renewcommand{\perp}{h}
\newcommand{\review}[1]{\textcolor{black}{#1}}

\def\volumeyear{2025}

\begin{document}

\runningheads{Huang J, van Reeuwijk M}{Multi-scale flow analysis for scale-aware urban-canopy models}

\title{Multi-scale flow analysis for scale-aware urban-canopy models}

\author{Jingzi Huang$^*$, Maarten van Reeuwijk}

\address{Department of Civil and Environmental Engineering, Imperial College London, London SW7 2AZ, UK}

\corraddr{jingzi.huang17@imperial.ac.uk}

\begin{abstract}
\begin{singlespace}
As Numerical Weather Prediction (NWP) models approach hectometric resolution, they increasingly operate in a regime where urban heterogeneity is only partially resolved and the assumptions underlying conventional urban canopy models (UCMs) become questionable. To address this scale gap, we apply a multi-scale coarse-graining framework (van Reeuwijk and Huang 2025, Boundary-Layer Meteorology) to building-resolving Large-Eddy Simulations (LES) of the University of Bristol campus, a realistic heterogeneous environment.
Two related morphologies are considered: an original layout containing large open-space contrasts and a modified configuration with these regions infilled. By systematically filtering the LES fields, we quantify how flow heterogeneity evolves with resolution and identify a characteristic urban length scale $\ell$ at which resolved and unresolved variability are comparable. This scale is strongly morphology-dependent, with $\ell \approx 256$ m for the original layout and $\ell \approx 64$ m for the modified case, demonstrating that neighbourhood-scale organisation can remain dynamically important at resolutions relevant to next-generation NWP.
Using this framework, we perform an \emph{a priori} assessment of distributed drag and turbulent-stress parameterisations. The results show that parameterisations derived from idealised geometries perform reasonably well only at sufficiently coarse resolutions ($L \gtrsim \ell$), where horizontal transport is negligible and the flow appears approximately homogeneous. At finer resolutions, their fidelity degrades rapidly due to increasing heterogeneity and filter-to-filter variability in morphology. These limitations are more pronounced in realistic layouts than in idealised cuboid arrays.
Overall, the results highlight that the applicability of urban parameterisations depends critically on the relationship between model resolution and a morphology-dependent heterogeneity scale. The framework provides a systematic route to diagnose this scale and to guide the development of scale-aware urban canopy models for high-resolution NWP.
\end{singlespace}

\vspace{1.0em}
\end{abstract}

\received{<details>}
\revised{<details>}
\accepted{<details>}

\maketitle

\section{Introduction}
Over recent decades, the resolution of Numerical Weather Prediction (NWP) models has increased to kilometre-scales or even hectometric scales, enabling the resolutions of most key atmospheric processes \citep{Bryan2003, Baldauf2011, Tang2013}. Nevertheless, it remains challenging for NWP models to capture the key processes within the urban canopy layer (UCL) --- the lowest portion of the atmospheric boundary layer. The UCL is characterised by extreme surface heterogeneity arising from buildings, street canyons, vegetation, and neighbourhood-scale variations (typically at $O(100-500 \U{m})$). Consequently, urban canopy models (UCMs), a branch of NWP models designed to represent sub-grid processes within the UCL, are required to parameterise the interaction of the atmosphere with the urban land surface that cannot be resolved explicitly.

Different levels of UCMs have been developed, ranging from single-layer schemes \citep{Kusaka2001, Kanda2005,Lipson2023} based on bulk parameters such as roughness lengths, displacement heights and planar area indices, to multiple-layer schemes \citep{Martilli2002, Coceal2004} that classify the UCL into vertical sub-layers according to the distribution of building heights and vegetation, and more recently to refined vertically distributed schemes that represent the evolution of momentum and heat with height \citep{Sutzl2021, Lu2024}. However, whether they rely on bulk parameters in single-layer form or on vertically distributed source terms derived from horizontal spatial averages, \review{these models generally assume the grid cell is horizontally homogeneous, with the horizontal plane average taken to represent it}. This assumption increasingly breaks down as NWP grids approach $O$(1 km) and below, where neighbourhood-scale urban heterogeneity such as clusters of tall buildings, open squares, and variable street orientations strongly modulates momentum transport, turbulence generation, and scalar dispersion \citep{Stewart2012, Masson2020}. Hectometric NWP models help address this issue \review{by partially resolving key physical processes, particularly boundary-layer turbulence}, which improves predictions of near-surface winds and turbulence kinetic energy in urban areas by up to $30\%$ compared to coarser models \citep{Lean2024}. However, at these scales the urban surface enters the so-called \emph{building grey zone}, in which large buildings and the flow around them are partially resolved on the grid while smaller-scale variability is not, so that surface parametrisations become more complex rather than simpler and horizontal exchanges between adjacent grid cells become first-order \citep{Lean2024}. To ultimately resolve this problem, it becomes necessary to relax the UCM assumption of horizontal homogeneity and to develop scale-aware models which are adjusted according to changes in model resolution \citep{Barlow2014}. Correspondingly, an analysis framework is required to systematically develop and evaluate such models across the scales.

To address these cross-scale challenges, the UK ASSURE (Across-Scale processeS in URban Environments) project, which is funded by Natural Environment Research Council (NERC), provides an integrated framework combining field observations, high-fidelity numerical simulations, and controlled wind-tunnel experiments. Focusing on the city of Bristol, characterised by complex topography and diverse urban morphology, ASSURE aims to determine which urban processes must be explicitly resolved and which require parameterisation in next-generation models. The effort includes integrating city-scale measurements \citep{Ludwig2025}, multi-source tracer experiments and DNS \citep{Auerswald2024, Clements2024}, building-resolving LES \citep{Coburn2022, MVR2025}, and atmospheric-boundary-layer (ABL) wind tunnel experiments \citep{Ding2023} at the Surrey Environmental Flow Laboratory (Enflo). The project provides a systematic basis for quantifying the evolution of urban flow and transport across scales.

As part of this effort, \citet{MVR2025} developed a multi-scale analysis framework which bridges the scale gap between the high-resolution LES and spatial average-based UCMs and therefore, provides a systematic way to examine how urban flow statistics and model performance evolve with resolution. Relying on a coarse-graining method applied to building‑resolving simulations, this framework is able to identify a characteristic urban length scale of a morphology at which the resolved and unresolved patterns are equally important. Moreover, the framework enables an \emph{a priori} diagnosis of urban drag and turbulence parameterisations using only morphological input. However, like many fundamental urban studies \citep{Macdonald1998, Coceal2004, Xie2008}, \citet{MVR2025} applied the framework on an idealised geometry. Although idealised geometries provide fundamental insights, they lack the heterogeneity and topographic complexity of real cities. Even within idealised cube and cuboid arrays, \citet{Blunn2022} found that fundamental closure parameters such as the von K\'arm\'an coefficient vary between approximately $0.20$ and $0.51$ across geometries, and that the mixing-layer analogy borrowed from vegetation canopies does not transfer to urban canopies because shear is local to individual buildings rather than dominated by a single canopy-top scale. Studies on realistic urban geometries have shown that the effective drag and mixing length are likewise not governed by a single length scale, but by a superposition of scales arising from individual buildings, street canyons, and district-level clustering \citep{Cheng2023}. Such findings highlight the limitations of extrapolating idealised parameterisations to real-world conditions and underscore the need to incorporate realistic geometry into multi-scale analyses \citep{Kusaka2024}.

The purpose of this study is to apply the multi-scale framework of \citet{MVR2025} to a realistic urban environment and investigate how urban heterogeneity evolves with scale and how sub-grid variability emerges as resolution is reduced on realistic morphology. Specifically, we aim (i) to investigate the characteristic urban length scale inferred for complex real morphologies and (ii) to assess, in an \emph{a priori} sense, the performance of UCMs over realistic morphologies, such as distributed drag and turbulence parameterisations. This is particularly important because, in contrast to uniform idealised layouts, the performance of such parameterisations is expected to vary more strongly across length scales in realistic, highly non-uniform configurations. Ultimately, this work offers new insights into the resolution-aware representation of realistic urban processes for NWP models. To address these questions, high-resolution LES are performed for the University of Bristol campus --- one of the main focus sites of the ASSURE project, which is characterised by strong heterogeneity in building shapes, a wide range of orientations and large variations in building height. 

The paper is organised as follows: \S \ref{sec: framework} reviews the multi-scale framework developed in \citet{MVR2025}. \S \ref{sec: simulation detail} details the simulations and the two realistic morphology layouts. Results are presented in \S \ref{sec: results}, comprising both a conventional plane-averaged analysis and the novel multi-scale analysis. Conclusions are drawn in \S \ref{sec: conclusions}.

\section{Multi-resolution planar-averaging framework on urban canopy flows} \label{sec: framework}
In this section, we review the computationally efficient multi-resolution planar-averaging framework of \citet{MVR2025}. The framework uses convolution filters to derive coarse-grained fields from a building-resolved high-resolution field. This approach provides a systematic means to implement the multi-scale analysis on the urban canopy. By applying the framework, we then introduce the planar-averaged drag and kinematic stresses, and their parameterisations, which will later be examined through an \emph{a priori} analysis.

\subsection{Planar averaging with convolution filter}
\begin{figure*}
    \centering
    \includegraphics[width = 14cm]{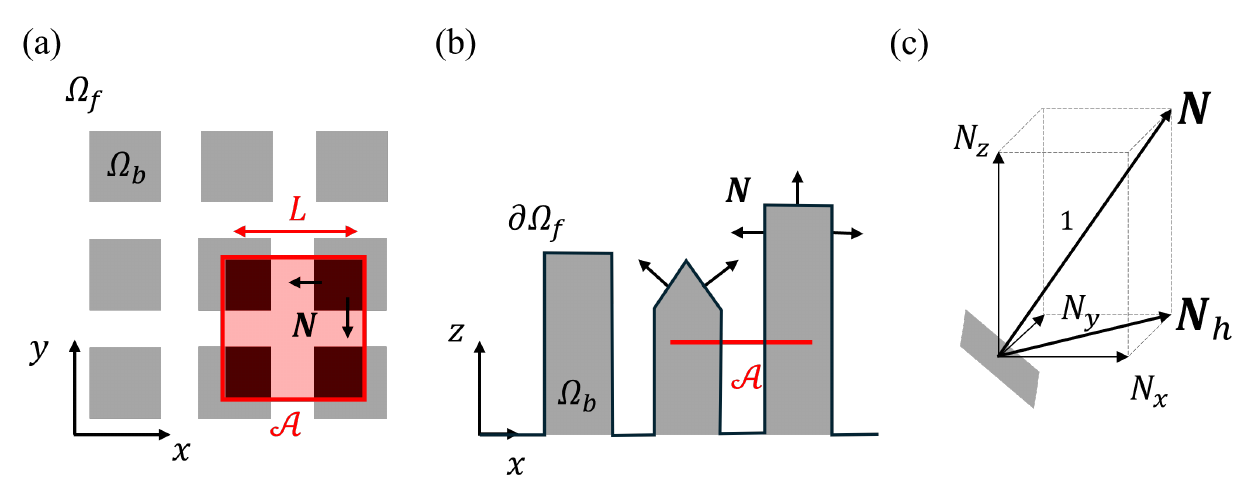}
    \caption{Definition sketch of domain and filter $\mathcal A$. (a) plan view, (b) elevation view. The domain of $\Omega$ is comprised of a fluid subdomain $\Omega_f$ (in white), a solid subdomain $\Omega_b$ (in grey), and a fluid-solid interface $\partial \Omega_f$ (solid black lines). The building-surface 3-D normal vectors $\vec N$ point into the fluid domain, and the decomposition is shown in (c). The 2-D square filter $\mathcal{A}$ with averaging length $L$ is shown in red.}
    \label{fig:sketch}
\end{figure*}

Consider a 3-D urban domain of interest $\Omega$ that is divided into a fluid subdomain $\Omega_f$ and a solid subdomain $\Omega_b$, separated by the fluid-solid interface $\partial \Omega_f$ (Figure \ref{fig:sketch}).
The coarse-graining method takes the \emph{superficial} planar average of an arbitrary field $\varphi$ over a square horizontal filter of length scale $L$, as illustrated in Fig. \ref{fig:sketch}(a, b).
Mathematically, this can be expressed as a convolution
\begin{equation} \label{eq:convolution}
 \sav{\varphi}_L(\vec x) =  \int_{\Omega_f(z)} \mathcal A(\vec x_\perp - \vec y_\perp)   \varphi(\vec y_\perp, z) \, \d \vec y_\perp \, ,
\end{equation}
where $\vec{x} = [\vec{x}_\perp; z]$, with $\vec{x}_\perp = [x, y]^T$ representing the streamwise and spanwise directions, and $z$ the vertical coordinate. The filter $\mathcal A(\vec x_h)$ is square and symmetric, centred at $\vec x_h$. The filter has value $L^{-2}$ inside the kernel but is $0$ elsewhere, ensuring the filter is normalised $\int \mathcal{A}(\vec{x}_h) \d \vec{x}_h = 1$. 
The length $L$ represents the averaging scale: the larger the value of $L$, the coarser the resulting field $\sav{\varphi}_L$. For simplicity, the subscript can be omitted when referring to a general concept if the averaging length $L$ is not emphasised, e.g., $\sav{\varphi}(\vec x)$. Particularly, when the averaging length $L$ is much larger than the characteristic urban length scale $\ell$, i.e., $L/\ell \gg 1$ or $L = \infty$, the quantity $\sav{\varphi}_{\infty}$ is identical to the conventional superficial plane-average $\sav{\varphi}_{xy}$, which is typically defined as \citep{Coceal2006, Xie2008, Giometto2016, Nazarian2020, Sutzl2020}
\begin{equation} \label{eq:slabav}
    \sav{\varphi}_{xy}(z) = \frac{1}{A_T} \int_{\Omega_f(z)} \varphi(\vec x) \, \d \vec x_\perp \, ,
\end{equation}
where $A_T$ is the total area of the horizontal plane.

\subsection{Planar-averaged stresses}

In this section, we show the exact form of the equations that need to be solved by non-building resolving NWP models in order to represent the effect of buildings. For simplicity, we restrict ourselves to a neutral case in which the $x$-direction is aligned with the mean wind, described with a large-scale forcing $f(\vec x)$. The time-averaged velocity and kinematic pressure are denoted by $\overline{u}_i$ and $\overline{p}$, respectively, with $\overline{u}$ representing the streamwise ($x$) velocity component. Using the coarse-graining formalism, the Reynolds-averaged momentum equation in the streamwise direction is given by \citep{MVR2025}:
\begin{equation} \label{eq:momentum}
  \frac{\partial \sav{\overline{u}_i}\sav{\overline u}}
  {\partial x_i}
  + \frac{\partial \sav{\overline p}}{\partial x} 
   = \sav{f} - \sav{f_D}
+ \frac{\partial \sav{\tau_i}}{\partial x_i},
\end{equation}
where
\begin{equation}
 \label{eq:slab_drag}
    \sav{f_{D}}(\vec x) \equiv 
     \underbrace{-\oint_{\partial \Omega_f}\mathcal A \overline p \frac{  N_x}{|\vec N_\perp|}\d s}_{\sav{f_{D,p}}} + \underbrace{\oint_{\partial \Omega_f} \mathcal A \nu \frac{\partial \overline u}{\partial x_j}  \frac{ N_j}{|\vec N_\perp|} \d s}_{\sav{f_{D,s}}} \, , 
\end{equation}
is the distributed drag term, comprised of a \review{form drag $\sav{f_{D,p}}$} associated with pressure, and a skin drag $\sav{f_{D,s}}$ component associated with viscosity. This equation also indicates that above the canopy layer, the drag is zero due to the absence of the building.
The total subfilter stress in Eq. \eqref{eq:momentum} is given by
\begin{equation}
\label{eq: tau_i}
     \sav{\tau_i}(\vec x)  \equiv
     \sav{\overline{u}_i}\sav{\overline u}- 
    \sav{\overline{u}_i \overline u} 
    - \sav{\overline{u^\prime_i u^\prime}} \, .
\end{equation}
Both the distributed drag Eq. \eqref{eq:slab_drag} and the subfilter stress Eq. \eqref{eq: tau_i} are terms that need to be parameterised in NWP models in order to reproduce the flow in the urban canopy layer. Note that both $\sav{f_D}$ and $\sav{\tau_i}$ terms are functions that depend on space $\vec x$ and also the filter's averaging length $L$.

It is useful to express the volumetric forcing and drag as the cumulative stresses from height $z$ upwards:
\begin{align}
    \tau_{f;L}(\vec x_\perp, z) & \equiv \int_z^h \sav{f}_L(\vec x_\perp, z') \d z' \, , \\ 
    \label{eq:tau_D}
    \tau_{D;L}(\vec x_\perp, z) & \equiv \int_z^h \sav{f_{D}}_L(\vec x_\perp, z') \d z' \, .
\end{align}

A main conclusion in \citet{MVR2025} states how the momentum balance Eq. \eqref{eq:momentum} can be simplified under multi resolutions: in the conventional horizontal plane-averaged sense, it is homogeneous in $x$ and $y$ directions, i.e., the averaging length is significantly large $L = \infty$, the horizontal components in Eq. \eqref{eq:momentum} vanish, when rewriting the equation by integrating it over the height, we obtain the following balance:
\begin{equation} \label{eq: int_force_balance}
  \sav{\tau_z}_\infty(z) = \tau_{f;\infty}(z) - \tau_{D;\infty}(z) \, ,
\end{equation}
which illustrates that the vertical transport of the streamwise velocity $\sav{\tau_z}_\infty$ is balanced by the wind-driven force and the building-induced drag. At the ground surface $z = 0$, Eq. \eqref{eq: int_force_balance} reduces to $\tau_{f; \infty}(0) = \tau_{D; \infty}(0) \equiv \tau_0$, where $\tau_0$ is the kinematic surface shear stress \citep{pope_2000}. 

Under the periodic boundary condition and applying triple decomposition \citep{Raupach1982, Finnigan2000b, Nepf2012}, the stress $\sav{\tau_z}_\infty$ can be further rewritten as
\begin{equation} \label{eq: tau_z_superposition}
\sav{\tau_z}_\infty =  - 
\sav{\overline{w} \, \overline u}_\infty 
- \sav{\overline{w^\prime u^\prime}}_\infty  = - \sav{\overline w^{\prime\prime} \overline u^{\prime\prime}}_\infty
-\sav{\overline{w^\prime u^\prime}}_\infty \, ,
\end{equation}
where $\sav{\overline w^{\prime\prime} \overline u^{\prime\prime}}_\infty(z)$ is the dispersive term, representing the deviation of the velocity from its plane average. 


\subsection {Parametrisations of drag and turbulence stresses}
Both the drag and kinematic stress terms are not resolved in NWP models and require parameterisations \citep{Raupach1982, pope_2000}. \citet{MVR2025} proposed their local parametrisations:
\begin{align}
\label{eq:3D-para}
    \frac{\tau_{D; L}}{\tau_{0;L}} &= A \zeta_L^3 -B  \zeta_L ^2 + (1-A-B)  \zeta_L \, ,   \\
\label{eq:tau_z_3d}
    \frac{\sav{\tau_z}_L}{\tau_{0;L}} &= 1 - A \zeta_L^3 +B  \zeta_L ^2 - (1-A-B)  \zeta_L \, ,
\end{align}
where $\sav{\tau_z}_L$ is the vertical kinematic stress within the filter. The parametrisations only depend on the initial conditions and morphology layout, represented by the cumulative normalised frontal area of the buildings over the horizontal plane $\zeta(z)$ 
\begin{equation}
 \zeta(z)=A_F^{-1}\int_z^{h_{\max}} L_b(z')\, \d z' \, , \quad A_F= \int_0^{h_{\max}} L_b(z')\, \d z' \, ,
\end{equation}
where $L_b$ is the frontal building width, $A_F$ frontal building area, and $h_{\max}$ represents the maximum building height. Correspondingly, $\zeta_L(\vec x)$ is the frontal area index covered by the filter, and it becomes $\zeta$ at $L = \infty$. We define the planar-averaged surface shear stress $\tau_{0;L} \equiv \tau_{D;L}(z = 0)$.
\citet{Sutzl2020} identified the parametrisation constant $A =1.88$ and $B = 3.89$ for $\lambda_p= 0.45$, $\lambda_f= 0.22$, but also noted that they are dependent on different morphological parameters. This morphology-dependence is consistent with broader evidence that closure parameters in urban canopy models are not universal: \citet{Blunn2022}, for instance, reported that the von K\'arm\'an coefficient varies by roughly a factor of 2.5 across idealised cube and cuboid arrays alone, with no clear systematic relationship to packing density.

It is important to note that the parameterisations Eq. \eqref{eq:3D-para} and Eq. \eqref{eq:tau_z_3d} are originally concluded from the plane-average balance Eq. \eqref{eq: int_force_balance}; therefore, they remain most accurate at large $L$ that horizontal homogeneity assumption can be made, i.e., $L \geq \ell$, but deteriorate as resolution increases, due to the assumption breaks down and the horizontal transport effects become significant.
Moreover, the turbulence parameterisation Eq. \eqref{eq:tau_z_3d} additionally assumes a constant stress canopy layer \citep{MVR2025}, which further limits its fidelity as the resolution increases. 

For consistency with earlier studies, we use the conventional notation $\sav{\varphi}_{xy}$ instead of $\sav{\varphi}_{\infty}$, and omit the infinity symbol for plane-averaged quantities when $L=\infty$, e.g. $\tau_{f; \infty} \equiv \tau_f, \tau_{D; \infty} \equiv \tau_D$. 
Table \ref{tab:vars} summarises the variables and symbols introduced in this section. The details of calculating $\sav{f_D}$ and $\zeta_L$ within an arbitrary filter require converting facet data into the Cartesian field, which are summarised in the Appendix.

\begin{table*}
    \centering
    \caption{\review{List of variables and symbols used in the following multi-scale analysis. The third column denotes short notation, when the averaging length $L$ does not matter. The fourth column corresponds to notation for $L = \infty$.}}

    \begin{tabular}{cccc}
        Name & Variable & Short notation& $L = \infty$\\
        Planar-averaged drag with averaging length $L$ &$\sav{f_D}_L(\vec{x})$ &  $\sav{f_D}$ &  $\sav{f_D}_{xy}(z)$\\
        Planar-averaged volumetric forcing with averaging length $L$ &$\sav{f}_L(\vec{x})$   & $\sav{f}$ & $\sav{f}_{xy}(z)$ \\
        Planar-averaged stress with averaging length $L$&$\sav{\tau_z}_L(\vec{x})$ &  $\sav{\tau_z}$&  $\sav{\tau_z}_{xy}(z)$\\
        Vertical integral of $\sav{f_D}_L$&$\tau_{D;L}(\vec{x})$ &  & $\tau_{D}(z)$\\ 
        Vertical integral of $\sav{f}_L$&$\tau_{f;L}(\vec{x})$ &  & $\tau_{f}(z)$\\
        Kinematic surface shear stress with averaging length $L$&$\tau_{0;L}(\vec{x}_h)$ & & $\tau_{0} \equiv \tau_{D}(0) = \tau_{f}(0)$\\
    \end{tabular}
    \label{tab:vars}
\end{table*}

\section{Simulation details} \label{sec: simulation detail}
\begin{figure*}
    \centering
    \includegraphics[width=16cm]{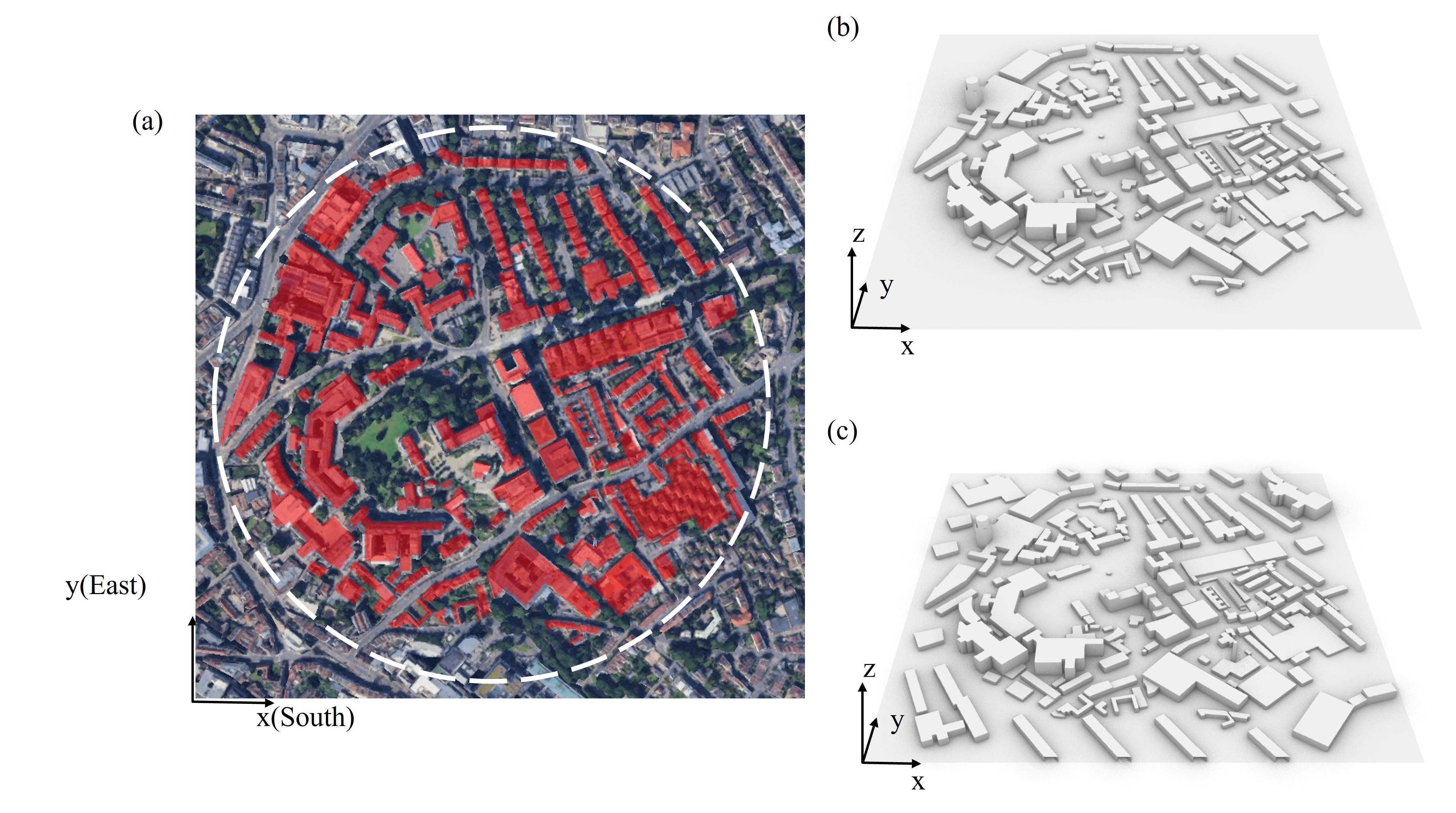}
    \caption{(a) A satellite map of the Bristol campus from Google Maps, with the footprints of the buildings, and \review{a dashed circle highlights the circular building configuration occupying the central area}. (b) The morphology of the `circular' case (CC) directly extracted from the map. (c) The morphology of the `square' case (SC), filling the corners in (b) with additional buildings.}
    \label{fig:Bristol_campus}
\end{figure*}
Figure \ref{fig:Bristol_campus} presents the urban morphology used in the simulations. The reference site is the campus of the University of Bristol (Fig. \ref{fig:Bristol_campus}a), one of the focus sites of the ASSURE project. Although the campus buildings naturally cluster around the main quadrangle, the precise circular footprint of the simulation domain is not a feature of the campus itself but is inherited from the corresponding wind-tunnel model deployed at the Surrey Environmental Flow Laboratory (EnFlo) \citep{Bi2025}. In wind-tunnel practice, an urban patch is constructed as a disc mounted on a turntable so that the incoming wind direction can be varied freely without re-machining the model; the disc therefore defines the buildable region of the experiment. The present LES geometry adopts the same disc cutout from the campus at $1:1$ scale (Fig. \ref{fig:Bristol_campus}b). When this disc is placed inside the square periodic computational patch ($L_x \times L_y \times L_z = 800\,\U{m} \times 800\,\U{m} \times 300\,\U{m}$), the four corners of the patch --- together about one quarter of $A_T$ --- remain as open space surrounding the buildings. We hereafter refer to this geometry as the \emph{disc configuration} of the campus, denoted CC (`circular case'). Beyond its experimental provenance, the disc configuration is well-suited to the present multi-scale analysis: the contrast between the densely-built disc and the surrounding open corners generates heterogeneity at the neighbourhood scale, complementing the building-scale variability inherent to any realistic urban morphology.

The buildings within the disc exhibit diverse shapes and orientations, with a mean building height of $h_m = 12$ m. The tallest building is a tower in the northeast, reaching $h_{\max} = 63$ m. The plan-area index and frontal area index are $\lambda_p \equiv A_p/A_T = 0.24$ and $\lambda_f \equiv A_F/A_T = 0.12$, respectively, where $A_p$ is the total building plan area, $A_F$ the frontal area, and $A_T = L_x L_y$ the horizontal domain area. These values reflect a layout with relatively extensive horizontal coverage (wide rooftops) but limited vertical development.

\review{The simulations are performed under periodic boundary conditions, and two morphology configurations are considered (Fig. \ref{fig:Bristol_campus}b,c). The first is the disc configuration described above, hereafter the `circular' case (CC). The second is a numerical modification, the `square' case (SC), in which the four open corners surrounding the disc are infilled with additional buildings sampled from the campus. This raises the plan-area index from $\lambda_p = 0.24$ to $0.31$ and the frontal area index from $\lambda_f = 0.12$ to $0.15$, while leaving the mean building height essentially unchanged. SC has no wind-tunnel counterpart; it is constructed numerically to suppress the neighbourhood-scale heterogeneity introduced by the open corners, leaving variability primarily at the building scale. Together, the two simulations enable a direct comparison between a disc geometry surrounded by a void and a fully built-up periodic patch.}

The simulations of urban flow over both morphologies were carried out using Large-Eddy Simulation (LES) with the open-source code uDALES \citep{Suter2022, Owens2024}. Solid boundaries were represented by the immersed boundary method (IBM), with near-wall dynamics parameterised by logarithmic wall functions \citep{Uno1995, Suter2022}. The turbulent eddy viscosity is calculated following the \citet{Vreman2004} subgrid model. The code employs a second-order central difference scheme on a staggered Arakawa C-grid for spatial discretisation and an explicit third-order Runge-Kutta scheme for time integration.

For both cases, the simulation domain size is $L_x \times L_y \times L_z = 800 \,\U m \times 800 \,\U m \times 300 \,\U m$ with a grid resolution of $N_x \times N_y \times N_z = 400 \, \times 400 \, \times 300$, corresponding to cell sizes of $\Delta x \times \Delta y \times \Delta z = 2 \,\U m \times 2 \,\U m \times 1 \,\U m$. Periodic boundary conditions are applied on the lateral boundaries, while a free-slip condition is imposed at the domain top. Atmospheric stability is neutral, with the wind driven by a constant streamwise body force $-\d P/\d x = 1.25 \times 10^{-5}$ kg s$^{-2}$ m$^{-2}$ in the streamwise ($x$) direction. Each simulation is run for 72\,000 s, and the last 60\,000 s of data are used for converged time-averaged statistics.

For the multi-scale analysis, eight different horizontal filter lengths were applied: $L = 4, 8, 16, 32,64,128,256$ and $512$ m. In addition, the conventional full-plane average over the entire horizontal domain ($800 \times 800$ m$^2$) was also included for comparison.

\section{Results} \label{sec: results}
\subsection{Plane-averaged statistics}
\begin{figure*}
    \centering
    \includegraphics{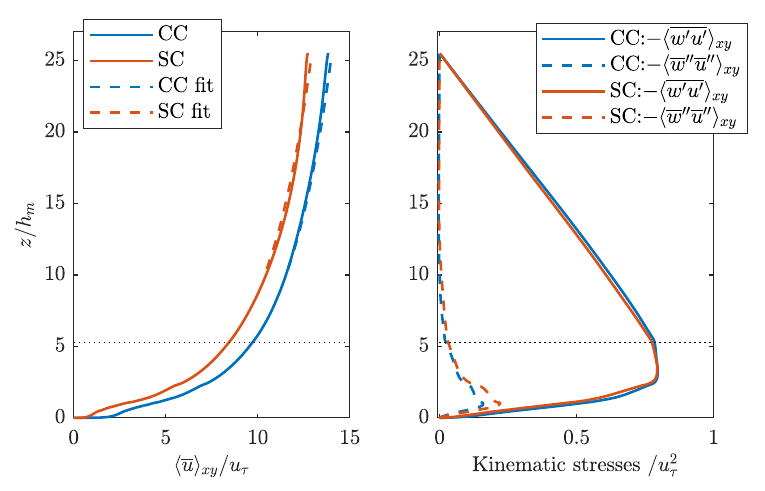}
    \caption{Superficially plane-averaged flow statistics for both cases. (a) streamwise velocity, overlaid with logarithmic profiles; (b) kinematic turbulent shear stress and dispersive stress. The height is normalised by the mean building height $h_m$ and the dashed horizontal line marks the maximum building height, i.e., the canopy top limit.}
    \label{Fig1}
\end{figure*}
We first examine the conventional plane-averaged statistics obtained from the simulations. Figure \ref{Fig1} presents the vertical profiles of the streamwise velocity and kinematic stresses for both cases. The profiles are normalised by friction velocity $u_\tau = \sqrt{\tau_0}$, where $\tau_0$ is the total surface kinematic stress. This stress can be predetermined from the imposed forcing via $\tau_0 = -\frac{\d P}{\d x} \frac{V_{air}}{A_T}$, where $V_{air}$ is the air volume inside the domain. \review{Note that the values of $\tau_0$ and $u_\tau$ in both cases coincide, as the difference in $V_{air}$ between the two layouts is negligible.}

As shown in Fig. \ref{Fig1}(a), CC has a higher velocity due to fewer building obstacles. Well above the canopy region, i.e., for $z \geq 2h_{\max}$, both velocity profiles fit the logarithmic law:
\begin{equation}
    \label{eq:loglaw}
    \spav{\overline{u}}(z) = \frac{u_\tau}{\kappa} \ln{\left( \frac{z-z_d}{z_0} \right)} \, ,
\end{equation}
where $\kappa=0.41$ is the Von Karman constant, $z_d$ is the displacement length and $z_0$ is the roughness length. The best-fit parameters are $z_{d} = 11.10 \, \U{m}, z_{0} = 1.08 \,\U{m}$ for CC and $z_{d} = 16.00 \,\U{m}, z_{0} = 1.64 \,\U{m}$ for SC. The increase in both displacement and roughness length in SC indicates that the additional buildings elevate the effective level of the mean flow and enhance surface roughness and drag.

Figure \ref{Fig1}(b) shows the vertical profiles of plane-averaged turbulent shear stress $\sav{\overline{w'u'}}_{xy}$ and dispersive stress $\sav{\overline{w}''\overline{u}''}_{xy}$. The dispersive stress $\sav{\overline{w}''\overline{u}''}_{xy}$ is significant within the canopy region, indicating strong flow heterogeneity in the mean flow induced by buildings. Above the canopy region, within the roughness sublayer, it gradually decays to zero (e.g., $z > 15 h_m$), indicating the transition to the inertial sublayer where the canopy influence completely disappears. This also confirms that the averaging time interval is sufficiently long to ensure statistical convergence. As expected, SC exhibits stronger dispersive stress within the canopy region than CC, due to the additional corner buildings that enhance variability.

Well above the canopy region, Fig. \ref{Fig1}(b) shows that the stress $\sav{\overline{w'u'}}_{xy}$ for both cases linearly decreases to zero with height, consistent with classical boundary layer theory. Their decreasing gradients are nearly identical, consistent with the fact that both cases are driven by the same pressure forcing. Interestingly, however, within the canopy region, the behaviour departs from previous studies of idealised cube arrays \citep{Coceal2004, Xie2008, Sutzl2020, MVR2025}, where $\sav{\overline{w'u'}}_{xy}$ typically increases monotonically from the ground to the canopy top. In the current realistic cases, $\sav{\overline{w'u'}}_{xy}$ first increases to a peak at approximately half the canopy height, but then slightly decreases before reaching the canopy top. \review{This is because this specific campus morphology: above half the canopy height, only the tall tower and a few isolated buildings remain, which generate much less drag than buildings below, therefore, their summation, the total kinematic stress $\sav{\tau_z}_{xy}$ peaks there (see discussion on Fig. \ref{Fig3} d,e), correspondingly, its main component $\sav{\overline{w'u'}}_{xy}$ also peaks here.}

\begin{figure*}
    \centering
    \includegraphics[width = 16cm]{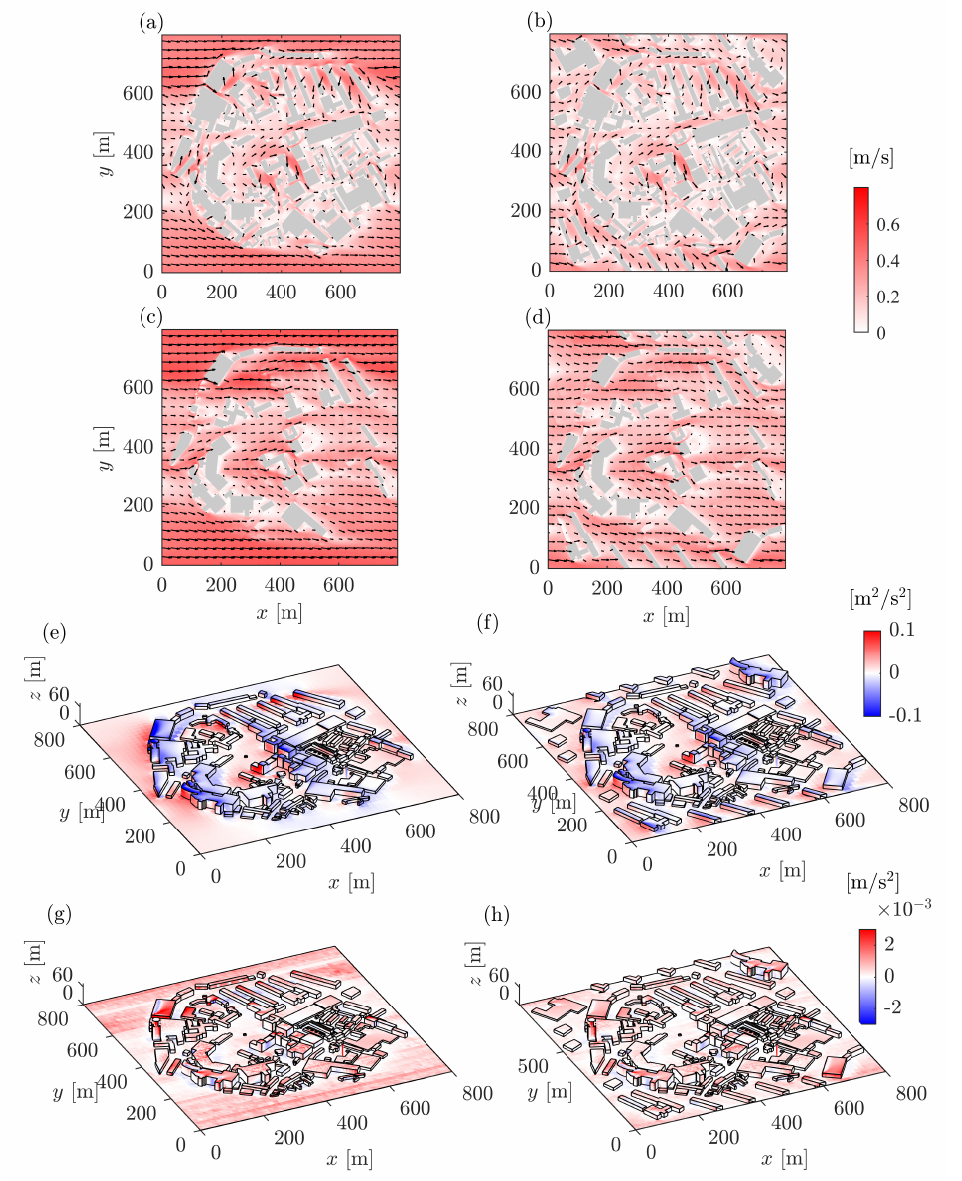}
    \caption{The plane view of the time-averaged wind speed ($u, v$) overlaid with the velocity vectors (a,b) at the pedestrian level ($z=1.5$ m), (c,d) at the mean building height ($z = h_m$), the grey area represents the buildings where there is no fluid. (e, f) The surface pressure of the solid phase. (g, h) The surface shear stress is induced by the streamwise velocity. The left column presents the CC, while the right column presents the SC.}
    \label{Fig2}
\end{figure*}

Figure \ref{Fig2} presents a comparison of both cases, with the left column corresponding to CC and the right to SC. The top two rows show the mean wind speed field, defined as $\sqrt{\overline{u}^2 + \overline{v}^2}$, at pedestrian level $z = 1.5 \U{m}$ (Fig. \ref{Fig2} a,b) and at the mean building height $z = h_m = 12 \U{m}$ (Fig. \ref{Fig2} c,d), respectively.
In Fig. \ref{Fig2}(a), channelling flows with high velocity can be observed at the top and bottom sides of the figure, reflecting the absence of building obstacles. This flow channelling is broken in the SC case  (Figs. \ref{Fig2}b,d). Near the circular building region (grey blocks), the velocity field becomes strongly heterogeneous, with clearly reduced wind speeds. At higher elevation ($z = h_m$), the building area decreases, and consequently the flow becomes more homogeneous, as shown in Fig. \ref{Fig2}(c). At this height, the wind direction remains largely aligned with the imposed streamwise forcing and is only locally deflected near the building.

Figures \ref{Fig2}(e,f) show the time-averaged building surface pressure field $\overline{p}$, obtained at the nearest grid cells from the building walls. Along the wind direction, the positive $x$-axis, positive pressure (dark red) can be observed on the upstream (frontal) faces of most buildings, while negative pressure can be observed in the wakes on the downstream (rear) sides (may not be visible in the present view). It is this pressure gradient that generates the form drag. In addition, negative pressure regions are clearly visible on building roofs, especially on the windward edges, which is consistent with flow boundary layer separation at the roof leading edges \citep{Atar2026}. A comparison between Figs. \ref{Fig2}(e) and \ref{Fig2}(f) highlight the shielding effect of the additional buildings: for example, along the upwind strip over $x \approx 150 - 200$ m, the roofs have less negative pressure in SC, as the incoming flow is decelerated by upstream obstacles.

Finally, Figs. \ref{Fig2}(g,h) show the distribution of shear stress, which is mainly contributed by the ground surface and building roofs. In CC, the shear stress is particularly strong along the ground-level channels, whereas in SC, due to additional building disruption, most shear stress is distributed to the roofs.
\begin{figure*}
    \centering
    \includegraphics[width = 15cm]{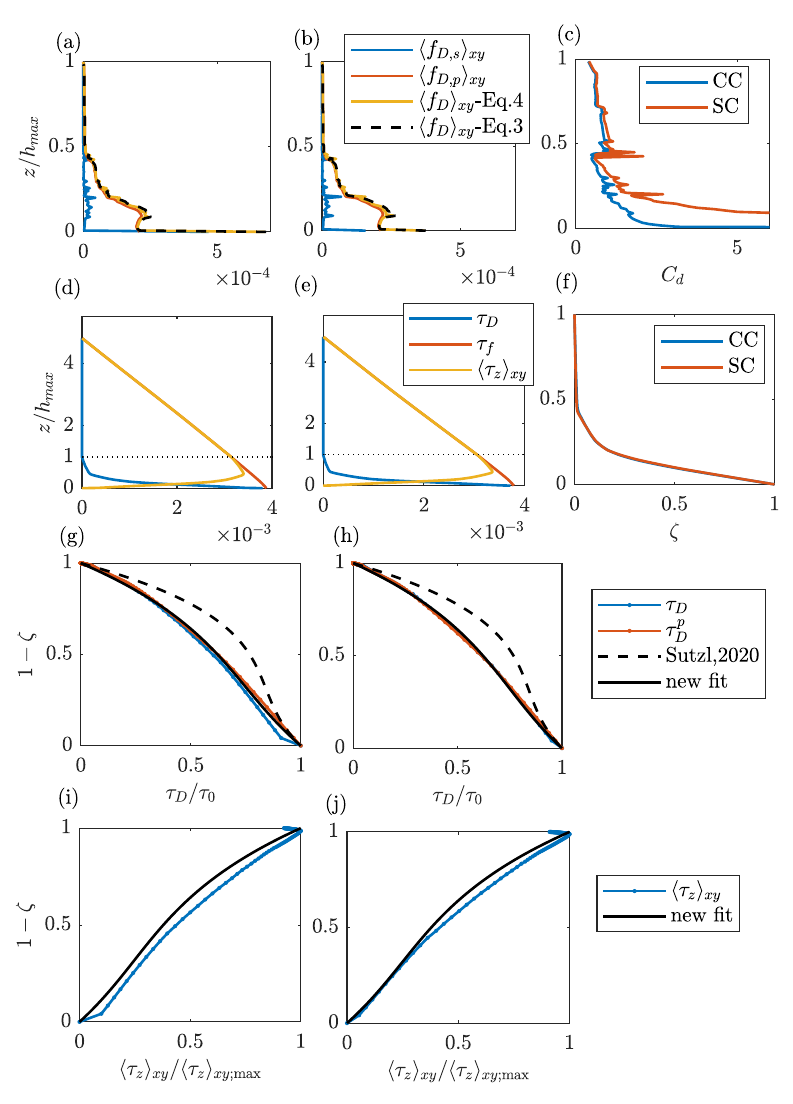}
    \caption{The vertical profile of plane-averaged (a, b) drag force and its components (unit m/s$^2$), (d, e) cumulative stresses (unit m$^2$/s$^2$), the height is normalised by the maximum building height marked as the horizontal dotted lines. (g, h) The cumulative drag stress $\tau_D$ against the normalised frontal area $\zeta$, overlaid with the parameterisation Eq. \eqref{eq:3D-para} at $L=\infty$; (i, j) The cumulative turbulent stress $\sav{\tau_z}$ against the normalised frontal area $\zeta$, overlaid with the parameterisation Eq. \eqref{eq:tau_z_3d} at $L=\infty$. The first column shows the profiles of CC, while the second column shows those of SC. (c) shows the vertical profile of the drag coefficient $C_d$; (f) shows the vertical profile of normalised frontal area $\zeta$ for both cases.}    
    \label{Fig3}
\end{figure*}

Figure \ref{Fig3} shows the profiles related to the drag. Figures \ref{Fig3}(a,b) present the vertical profiles of the plane-averaged drag force and its form and skin components for CC and SC, respectively. The total drag force, $\sav{f_D}_{xy}$, is obtained in two ways: first, from the momentum budget closure (Eq. \eqref{eq:momentum} with horizontal terms neglected in the plane-averaged perspective), and second, from the direct line integral (Eq. \eqref{eq:slab_drag}). The results indicate that in both cases, the two methods are in close agreement: the drag force peaks near the mean building height and diminishes to nearly zero above $z/h_{\max} > 0.5$ due to the absence of buildings. This is consistent with building normalised frontal area $\zeta$ in  Fig. \ref{Fig3}(f), where it clearly shows that there is very little building surface area above $z/h_{\max} > 0.5$.
As suggested by the kinematic stresses in Fig. \ref{Fig1}(b), the overall magnitude of plane-averaged $\sav{f_D}_{xy}$ is similar between the two cases, with SC slightly larger. However, notably, the skin drag at the ground level $z = 0$ in CC is significantly greater than in SC. This agrees with the ground shear stress distributions shown in Figs. \ref{Fig2}(g,h) and is quantitatively confirmed here.

Although the form drag $\sav{f_{D,p}}_{xy}$ dominates the total drag in both cases, ignoring surface shear stress (the skin drag) should be done carefully. Specifically, at $z/h_{\max} \approx 0.4$, shear stress contributes up to nearly half of the total drag. This can be attributed to the large footprint of the buildings, as indicated by the frontal-area index $\lambda_f$. Since drag components can only be determined directly from Eq. \eqref{eq:slab_drag}, hereafter, we will use Eq. \eqref{eq:slab_drag} to calculate the drag for further analysis.

Fig. \ref{Fig3}(c) presents the vertical profiles of a drag coefficient $C_d(z)$ \citep{Coceal2004, Santiago2010}, which relates directly to the distributed drag $\sav{f_D}_{xy}$ through
\begin{equation}  \label{eq: Cd}
    \sav{f_D}_{xy} =\frac{1}{2}  \frac{L_b}{A_T} C_d \sav{\overline{u}}_{xy}^2 \, .
\end{equation}
It is an alternative way of drag parameterisation from Eq. \eqref{eq:3D-para}, and this drag coefficient $C_d$ is widely adopted in engineering practice; therefore, it is useful to include it here.
The drag coefficient is very large near the ground due to the low plane-averaged velocities, which is consistent with \citet{Coceal2004}. The SC exhibits larger $C_d$ values than CC throughout the lower half of the canopy due to the additional buildings. In the upper half, the difference decreases due to the absence of the additional buildings. The drag coefficient obtained here is lower than that reported for idealised staggered arrays (between 2 and 3) by \citet{Coceal2004}. The building orientation is likely one of the reasons for this discrepancy: in current cases, varied building alignments deflect flow and reduce streamwise drag, whereas in the idealised cases of \citet{Coceal2004}, wind is always perpendicular to the cubes.

Figures \ref{Fig3}(d,e) show the plane-averaged cumulative stresses, which satisfy the balance in Eq. \eqref{eq: int_force_balance}. In line with the calculation of $\tau_0$, the surface kinematic stresses are nearly identical for both cases. Recall that $\sav{\tau_z}_{xy}$ represents the superposition of turbulent and dispersive stresses (Eq. \eqref{eq: tau_z_superposition}), already shown in Fig. \ref{Fig2}(b). This stress peaks near the mean building height and decreases linearly above the canopy. The cumulative drag $\tau_D$ decreases smoothly with height, but its slope sharpens at approximately $z/h_{\max} = 0.5$, coinciding with the sudden reduction in building area, consistent with the vertical profiles in Figs. \ref{Fig3}(a, b, f).

The drag parameterisation (Eq. \eqref{eq:3D-para}) in the plane-averaged sense is examined in Figs. \ref{Fig3}(g,h). These figures illustrate the evolution of $\tau_D$ with $1 - \zeta$ (where $1 - \zeta = 0$ corresponds to the ground), overlaid with the parameterisations. Note that the parameterisation in Eq. \eqref{eq:slab_drag} originally considered only the form drag and neglected the skin drag; therefore, both the full cumulative drag $\tau_D$ and the form-drag-only component $\tau_D^p$ are presented here. Typically, in terms of the vertical accumulation, the skin drag component is indeed negligible, as $\tau_D$ and $\tau_D^p$ appear very similar. However, it is worth noting that, near $1-\zeta = 0$ at ground, the profile of $\tau_D$ in CC exhibits a discontinuous reduction, whereas this does not occur in SC and any $\tau_D^p$ profiles. This reduction arises from the high ground shear stress in CC induced by the open spaces, an effect that extends up to the canopy top.
However, neither of the profiles matches the parameterisation of \citet{Sutzl2020} (dashed lines), which substantially overestimates drag. This discrepancy exists because the parameterisation constants depend on morphological parameters: the constants in \citet{Sutzl2020}, $A=1.88$, $B=3.89$, were derived from morphologies with larger plan- and frontal-area indices than in the present study. Given that both indices here are about one-third smaller, a reduced drag is expected. By refitting Eq. \eqref{eq:3D-para}, new constants $A = 0.89$ and $B = 1.82$ are obtained, shown as solid black lines in Figs. \ref{Fig3}(g,h). The coefficient of determination $R^2>99.8\%$ for both cases indicates good agreement.

Finally, Figs. \ref{Fig3}(i,j) evaluate the parameterisation of turbulent stress $\sav{\tau_z}_{xy}$ using Eq. \eqref{eq:tau_z_3d} with the newly fitted constants. The new formulation demonstrates good agreement in both cases. Here, the normalisation is slightly modified: instead of using the surface stress $\tau_0$, the maximum stress $\sav{\tau_z}_{xy;\max}$ is adopted to constrain the parameterisation within the unit interval. This follows the constant stress layer assumption \citep{Schlichting2017}, i.e. $\tau_f(z<h_{\max}) \approx \tau_0 \approx \sav{\tau_z}_{xy;\max}$, which is more accurate when the domain height is much greater than the canopy height. The normalised stress decreases monotonically with $1-\zeta$, except near $1-\zeta = 1$ (the upper half of the canopy), where minor deviations occur but remain close to the fitted curve. 

Overall, the revised parameterisation (Eq. \eqref{eq:tau_z_3d}) with the new constants predicts the turbulent stress within the canopy for the present morphologies reasonably well. 

\subsection{Multi-resolution analysis on heterogeneity}
\begin{figure*}
    \centering
    \includegraphics{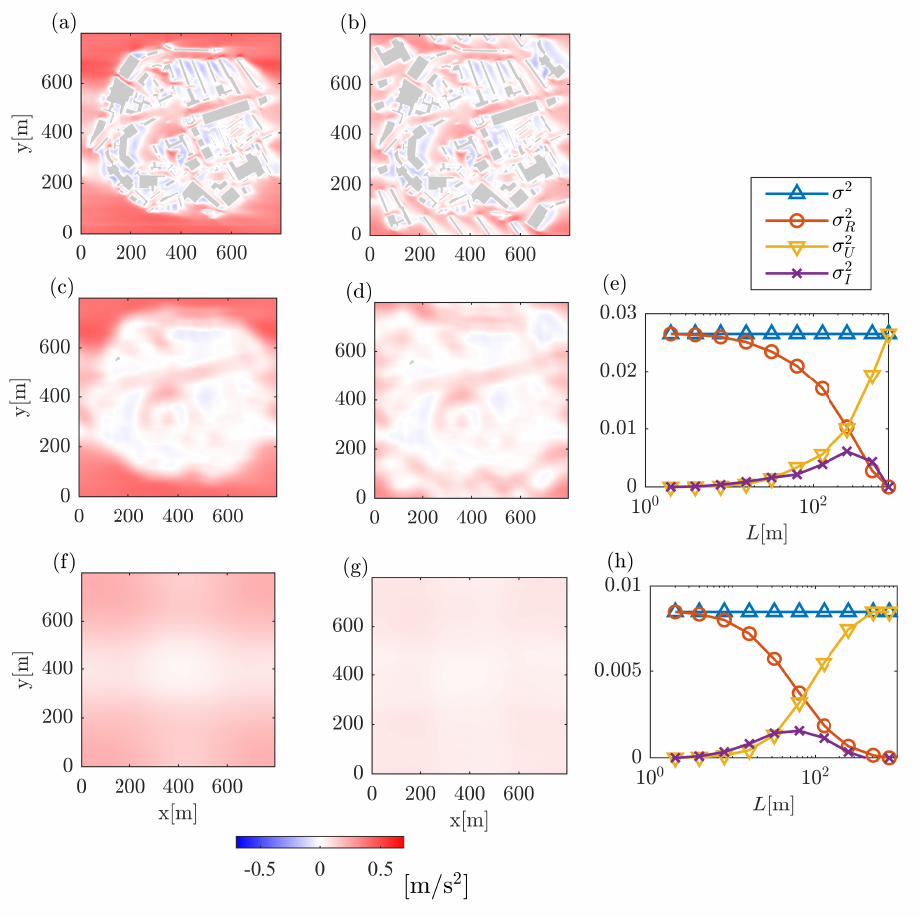}
    \caption{The streamwise velocity field $\overline{u}$ at various averaging lengths at the pedestrian level $z = 1.5 \U{m}$ for CC (the first column) and SC (the second column): (a, b) averaging lengths $L = 8\U{m}$, (c, d) $L = 64$ m,  (f, g) $L = 512$ m; (e, h) are the variances of the streamwise velocity at pedestrian level for CC and SC, respectively.}
    \label{Fig4}
\end{figure*}
The multi-resolution framework is applied to this realistic campus, although it has turned out to work very successfully on idealised geometry \citep{MVR2025}. The multi-resolution fields generated by the coarse-graining method are presented first. 
The first two columns in Fig. \ref{Fig4} show the coarse-grained streamwise velocity $\sav{\overline{u}}_L$ at pedestrian level ($z = 1.5 \, \U{m}$) for CC and SC, obtained with averaging lengths $L = 8 \, \U{m}, 64\, \U{m}$ and $512\, \U{m}$. The sequence clearly illustrates how the field evolves with the resolution. At the relatively high resolution $L = 8\, \U{m}$ (Figs. \ref{Fig4}a, b; note that the original resolution is $1 \U{m}$), the details of the velocity field are clearly retained, showing strong heterogeneity: individual buildings remain visible, and negative velocities appear in the wakes and street canyons. At $L = 64\, \U{m}$, (Figs. \ref{Fig4}c, d), the velocity fields become blurry and only large-scale features can be still identified, such as the long street extending from $x = 200 \, \U{m}, y = 400 \, \U{m}$ to $x =700\, \U{m}, y = 500 \, \U{m}$. At this resolution, most buildings disappear, and their footprint is represented by very low velocities. Interestingly, in CC (Fig. \ref{Fig4}(c)), these low-velocity regions clearly trace out the outline of the disc configuration, with the building-occupied cluster standing apart from the higher-velocity open corners.
At the coarsest resolution $L=512 \, \U{m}$ (Figs. \ref{Fig4}f, g), the velocity fields become nearly homogeneous and close to the entire plane-averaged values. Although no velocity details remain, a difference can still be observed: in CC (Fig. \ref{Fig4}f), the four corners exhibit higher velocities while the centre shows lower values; however, in SC, such a feature is not significant due to the additional buildings at the corners. It is worth noting that even the coarsest resolution here, $L=512 \, \U{m}$, is still finer than most NWP models, which indicates that the capacity of NWP models to capture heterogeneous details in the urban canopy region is still limited.

The spatial heterogeneity of the streamwise velocity $\overline u $ can be quantified by its variances \citep{Yu2023, MVR2025}. The plane-averaged variance $ \sigma^2 $ is defined as
\begin{equation}
    \sigma^2(z) \equiv \spav{(\overline u(\vec x_\perp, z) - \spav{\overline u})^2} \, ,
\end{equation}
which can be decomposed as
\begin{equation}
  \label{eq:variancedecomp}
  \sigma^2 = \sigma^2_{R} + \sigma^2_{U} + \sigma^2_{I} \, ,
\end{equation}
where
\begin{align}
\label{eq: sigma_U}
{\sigma^2_{U}}(z) & \equiv {\spav{(\overline u - \sav{\overline u}_L)^2}}, \\
\label{eq: sigma_R}
{\sigma^2_{R}}(z)& \equiv{\spav{(\sav{\overline u}_L - \spav{\overline u})^2}}, \\
{\sigma^2_{I}}(z) & \equiv {2 \spav{(\overline u - \sav{\overline u}_L)(\sav{\overline u}_L - \spav{\overline u})}} \, .
\end{align}
Here, $\sigma^2_{U}$ is the unresolved variance, representing the discrepancy between the local value (e.g., $\overline{u}$) and the planar-averaged value (e.g., $\sav{\overline u}_L$). A large unresolved variance typically indicates a significant difference between the original high-resolution field and the field at resolution $L$, suggesting that the field features at scale $L$ are largely unresolved and therefore require modelling. The component $\sigma^2_{R}$ is the resolved variance. A small resolved variance typically implies that the planar-averaged value at resolution $L$ is close to the average over the entire horizontal plane (e.g., $\spav{\overline{u}}$), again, indicating that the field at resolution $L$ is very coarse. A large resolved variance suggests that the field is well resolved.
The component ${\sigma^2_{I}}$ is (twice) the interaction covariance \citep{MVR2025}. Note that, by definition, the total variance $\sigma^2$ is independent of the averaging length $L$, but its components are.

Figure \ref{Fig4}(e) shows the decomposed variances for CC at pedestrian height varying with the averaging length $L$. As $L$ increases, the resolved variance $\sigma^2_R$ decreases while the unresolved variance $\sigma^2_U$ increases, consistent with the coarse-graining process in Fig. \ref{Fig4}(a,c,f). At high resolutions, the resolved variance dominates, indicating that the field is largely resolved and it is not necessary to use a model. Conversely, at low resolutions, the unresolved component dominates, indicating a model is needed to resolve the field. The resolution where the resolved and unresolved variance crossover indicates a scale where the resolved and unresolved parts are equally important; therefore, we refer to this length scale as the characteristic urban length $\ell$ of the morphology. This length scale provides a concept of the extent to which the field may be regarded as homogeneous. \review{For CC, $\ell \approx 256$ m, which is comparable to the radius of the circular patch and corresponds to a neighbourhood length scale. This scale indicates that the circularly arranged building cluster can be distinguished from the surrounding open space at a resolution of approximately $256$ m. At coarser resolutions, the field shows effectively homogeneous, in contrast, to capture building-scale structures, like buildings and streets, finer resolutions than this scale are required.}
By definition (Eq. \eqref{eq:variancedecomp}), the interaction term $\sigma^2_I$ is only significant when resolved and unresolved variances are both important, and indeed, its peak coincides with $\ell$.

Figure \ref{Fig4}(h) presents the same analysis for SC. The total variance is smaller ($\sigma^2 = 0.01$) than in CC ($\sigma^2 = 0.03$), reflecting the more homogeneous velocity distribution at the corners. A key difference is that in SC, as $L$ increases, $\sigma^2_R$ decreases more rapidly (and equivalently, $\sigma^2_U$ increases more rapidly) compared to CC; consequently, the curves intersect at a smaller length scale $\ell \approx 64$ m. This is because, technically, in CC, the corner values of $\sav{\overline u}_L$ remain relatively large as $L$ increases. As a result, the reduction of $\sigma_R^2$ and the growth of $\sigma_U^2$ with $L$ occur more slowly in CC. \review{This length scale is comparable to a characteristic building scale, indicating that heterogeneity in SC typically occurs at the building scale. At coarser resolutions, such variability is not resolved.}
\begin{figure*}
    \centering
    \includegraphics{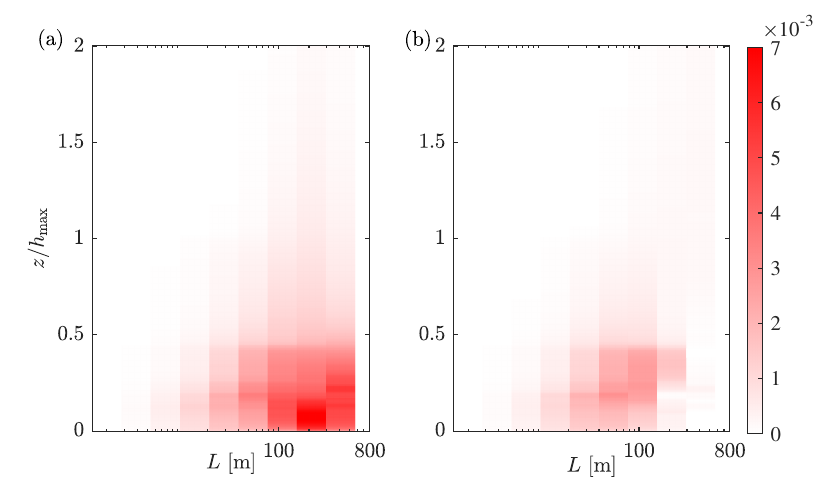}
    \caption{The contour plot of the interaction covariances of streamwise velocity against the averaging length $L$ and the height $z$, for CC (a) and SC (b), respectively.}
    \label{Fig6}
\end{figure*}

To have a full scope of characteristic length scale over the height, the peak of interaction covariance $\sigma^2_I$ is useful, as $\sigma^2_I$ indicates the correlation between resolved and unresolved components. Figure \ref{Fig6} shows $\sigma^2_I$ variation with both averaging length $L$ and height $z$, with height normalised by the maximum building height $h_{\max}$.
For both CC and SC, the interaction covariance is small near and above $h_{\max}$, where building effects on the flow are weak, flow tends to be uniform. While within the canopy region, $\sigma^2_I$ exhibits distinct peaks at different averaging lengths: around $L = 256-512$ m for CC, and around $L = 64-128$ m for SC. \review{This pattern echoes the variance-crossover analysis at pedestrian level (Fig.~\ref{Fig4}e,h) and extends it through the canopy depth. In both cases $\sigma^2_I$ is concentrated below $z/h_{\max} \approx 0.5$, where most building footprints reside, but the CC peak is roughly twice as strong as its SC counterpart and remains detectable up to $z \sim h_{\max}$, indicating that the disc-scale heterogeneity is felt across the full canopy. The SC peak is weaker, narrower in $L$, and confined to the lower canopy, consistent with a heterogeneity whose scale is set by individual building footprints rather than by any larger-scale organisation. Physically, the two cases differ in how unevenly the buildings are distributed within the periodic patch: in the disc configuration, the contrast between the densely-built disc and the open corners produces a heterogeneity whose dominant scale reflects the cluster radius; filling those corners (SC) removes the disc-versus-void contrast, leaving only the variability associated with individual buildings and street canyons. The diagnosed $\ell$ therefore depends not on packing density alone but on how the buildings are distributed across the domain --- a sensitivity that is generic to realistic urban morphologies, where building layouts are rarely statistically uniform.}

\review{Although the variance analysis above uses the streamwise velocity, the same characteristic length scale $\ell$ is recovered when other fields are used, e.g.  the turbulent kinetic energy field as demonstrated on idealised geometries by \citet{MVR2025}}.

%
\begin{figure*}
    \centering
    \includegraphics{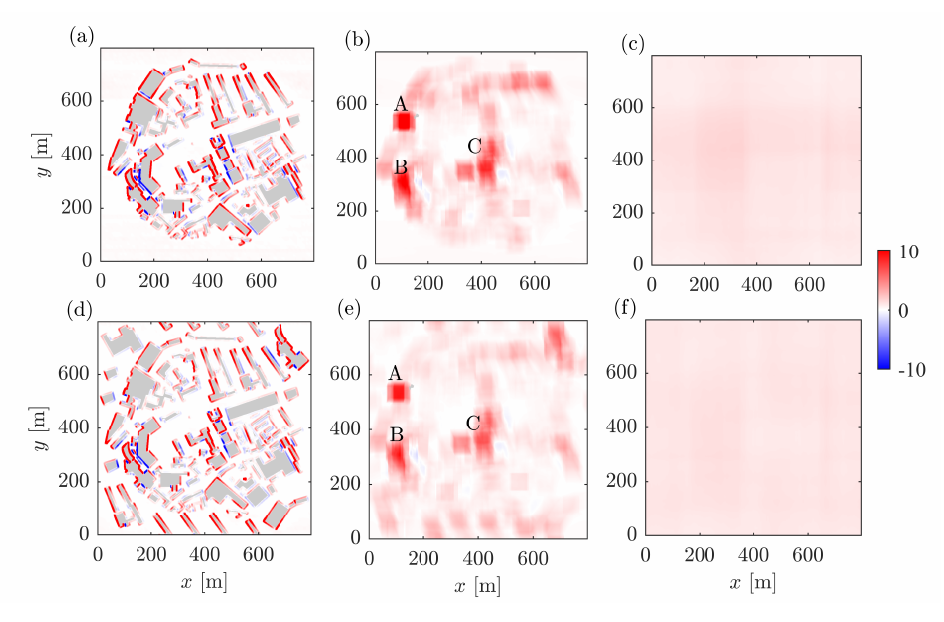}
    \caption{The coarse-grained kinematic surface stress $\tau_{0;L}$ for CC (top panel), and SC (bottom panel), respectively, with the averaging lengths (a, d) $L = 8$ m, (b, e) $L = 64$ m and (c, f) $L = 512$ m. Three high-drag zones are labelled in (b, e).}
    \label{Fig7}
\end{figure*}

To examine the drag on buildings, Fig. \ref{Fig7} presents the spatially averaged kinematic surface stress $\tau_{0;L}$ at three resolutions ($L = 8, 64, 512$ m) at the pedestrian level. As $\tau_{0;L} \equiv \tau_{D;L}(z=0)$ (Eq. \eqref{eq:tau_D}), it represents the filter-averaged integral drag force exerted on the buildings.
At high resolution $L = 8$ m (Figs. \ref{Fig7}a,d), $\tau_{0; L}$ clearly outlines the buildings, as the drag stress mostly occurs at the building surface. A faint red is also visible over the ground, which is due to the ground surface shear stress. This resolution provides a detailed view of the stress distribution around the building. For example, in the region $x < 200$ m, the dark red area indicates that the windward-facing buildings experience large drag, while negative drag stress can be observed in the wake region between locally dense building clusters. Note that, as an integral value, the magnitude of $\tau_{0;L}$ is correlated closely with the building height.

At an intermediate resolution of $L = 64$ m (Figs. \ref{Fig7}b,e), the drag patterns become more diffuse, and individual buildings are no longer visible, consistent with the behaviour seen in Fig. \ref{Fig4}. Instead, three distinct high-drag regions can be noticed (with dark red): zone A contains the tallest tower, whose large height enhances drag; zones B and C have relatively high building density and moderate building height, which also generate significant drag. While no strong negative drag areas are apparent at this resolution, very light blue patches appear downstream of zones B and C, suggesting wake effects in the leeward regions. Thus, this resolution, typically close to the characteristic length scale, is suitable for monitoring key areas of interest within the domain. 

At the coarsest resolution $L = 512$ m (Figs. \ref{Fig7}c,f), the heterogeneity is not visible. The surface stress $\tau_{0;L}$ becomes nearly homogeneous across the entire plane, with magnitudes converging to the global average $\tau_0$.
\begin{figure*}
    \centering
    \includegraphics{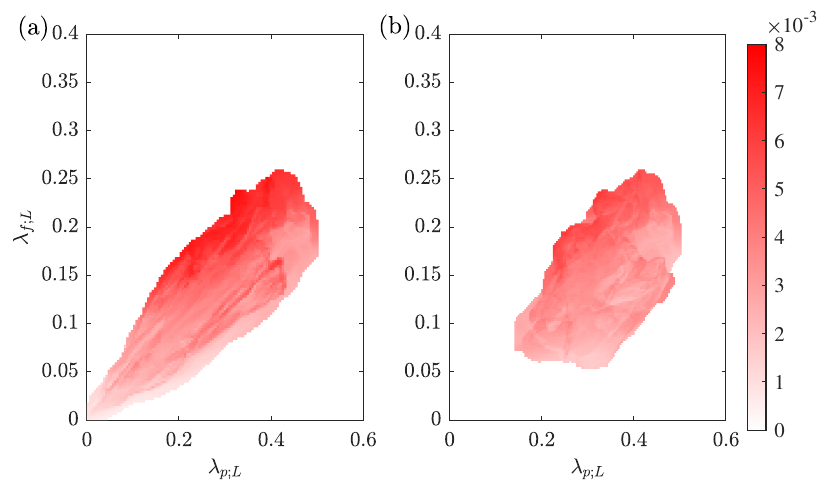}
    \caption{Spatial-averaged kinematic surface stress $\tau_{0; L}$ against the plane-area index $\lambda_{p; L}$ and front-area index $\lambda_{f; L}$ for (a) CC, (b) SC, respectively. \review{The data of averaging lengths $L = 256, 512$ m are used in both figures}.}
    \label{Fig12}
\end{figure*}

Since the kinematic surface stress $\tau_{0;L}$ represents the drag acting on buildings, and also serves as the normalisation factor in the parameterisation of Eq. \eqref{eq:3D-para}, it is useful to examine how it varies with morphological parameters.
Figure \ref{Fig12} shows two joint PDFs of $\tau_{0;L}$ against the local plane-area index $\lambda_{p;L}$ and the frontal-area index $\lambda_{f;L}$ for both cases. Here, we use the data at $L = 256$ m and $512$ m, as these fall within the valid range of the parameterisation Eq. \eqref{eq:3D-para}. Recall that this parameterisation is obtained based on a plane-averaged sense (Eq. \eqref{eq: int_force_balance}), in which only vertical transport is considered, and horizontal transport is assumed to be negligible.

The broad distribution of $\tau_{0;L}$ across $\lambda_{p;L}$ and $\lambda_{f;L}$ confirms that the constants in the parameterisation (Eq. \eqref{eq:3D-para}) indeed depend on the underlying morphological parameters. As expected, CC (Fig. \ref{Fig12}a) exhibits a wider spread in both $\lambda_{p;L}$ and $\lambda_{f;L}$ compared to SC (Fig. \ref{Fig12}b), with an additional `tail' extending toward the origin. This arises because, in CC, the filter can cover corner regions with very few or even no buildings, leading to very low $\lambda_{p;L}$ and $\lambda_{f;L}$ and correspondingly low surface stresses.


An important observation is that, in both morphologies, $\tau_{0;L}$ tends to increase when $\lambda_{f;L}$ increases, as larger building frontal area induces greater drag, or when $\lambda_{p;L}$ decreases, as reduced shielding enhances drag. This trend becomes particularly pronounced for $\lambda_{f;L} > 0.15$. Furthermore, the results indicate that $\tau_{0;L}$ can remain nearly constant if $\lambda_{p;L}$ and $\lambda_{f;L}$ vary simultaneously along a trajectory with an approximate $45^\circ$ slope, as the opposing effects of increased frontal area and reduced shielding balance each other.
\begin{figure*}
    \centering
    \includegraphics{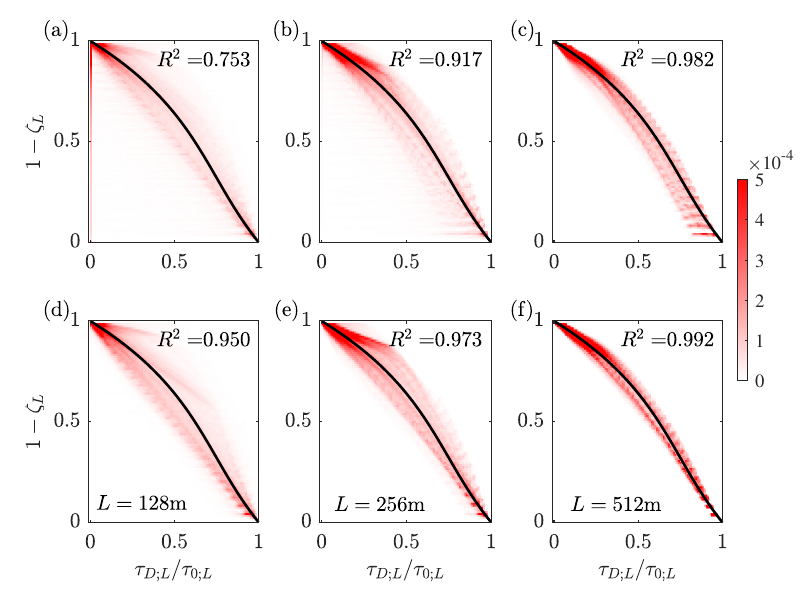}
    \caption{The distribution density of normalised local cumulative drag stress $\tau_{D; L}/\tau_{0;L}$ against the local scaled frontal area $\zeta_L$ at different averaging lengths. (a, d) $L =128$ m, (b, e) $L=256$ m, (c, f) $L= 512$ m, for CC (top panel) and SC (bottom panel), respectively. The colour represents the statistical probability density function value of the data appearing at each location. The solid line represents the parametrisation Eq. \eqref{eq:3D-para} with modified constants $A = 0.89, \, B = 1.82$. The coefficient of determination $R^2$ is labelled to show the performance of the parametrisation.}
    \label{Fig8}
\end{figure*}

Next, we perform an \emph{a priori} analysis at different averaging lengths $L$ to evaluate the local drag parameterisation in Eq. \eqref{eq:3D-para}. Figure \ref{Fig8} presents the distribution density of the local cumulative drag stress $\tau_{D;L}$ against the local scaled frontal area $\zeta_L$ at three resolutions ($L = 128$, $256$, and $512$ m), overlaid with the parameterisation (with modified constants) for both cases.

The parametrisation performs best in Fig. \ref{Fig8}(f): at low resolution $L = 512$ m in SC, with a high coefficient of determination ($R^2 = 0.998$). At this resolution, the data scatter is minimal because the fields are relatively homogeneous, as already shown in Fig. \ref{Fig7}(f). Thus, a parameterisation originally derived from a plane-averaged perspective remains satisfied. As the resolution increases (Figs. \ref{Fig8}e,d), the data scatter increases, and heterogeneity rises. At smaller filter lengths, horizontal transport terms in the momentum balance become important, while parameterisation only considering vertical terms becomes less accurate. At even higher resolutions (e.g., $L < 64$ m, not shown here), the parameterisation rapidly deteriorates. Indeed, when the resolution varies across the characteristic length, for SC, from $L=256$ m to $L = 64$ m, there is a significant increase in the resolved variance (Fig. \ref{Fig4}h), and thus more heterogeneity and details which is a challenge for parametrisation to capture. However, the parameterisation still works reasonably when the averaging filter length is greater than the characteristic length $\ell$ in this case.

In contrast, CC shows worse convergence. Even at the same low resolution ($L = 512$ m, Fig. \ref{Fig8}c), the data distribution is more widely spread from the bottom, indicating that the normalised cumulative drag $\tau_{D;L}/\tau_{0;L}$ is hard to converge. This divergence arises because the filter in CC can cover a wider range of morphological conditions, $\lambda_{p;L}$ and $\lambda_{f;L}$, because of the corners (see Fig. \ref{Fig12}a). Thus, consequently a wider range of $\tau_{D;L}$ and $\tau_{0;L}$ than SC. Since the constant in parameterisation Eq. \eqref{eq:3D-para} depends on morphology, a single set of constants cannot adequately capture such variability. 
As a result, the $R^2$ in CC (Fig. \ref{Fig8}c) is significantly lower than that in SC (Fig. \ref{Fig8}f).

Further increasing the resolution, Fig. \ref{Fig8}(a, b) present an even worse fit, either compared with Fig. \ref{Fig8}(c) in the same case or compared with Fig. \ref{Fig8}(d, e) at the same resolution, respectively. Notably, Fig. \ref{Fig8}(a) shows instances with $\tau_{D;L}/\tau_{0;L} \approx 0$, corresponding to filters that cover very few building areas thus almost no drag, for example, at open corner regions. Such cases hardly appear at larger $L$, where filters inevitably include more buildings. For CC, the resolution at which the parameterisation still performs reasonably is around $L = 256$ m, which aligns with its characteristic length scale.
\begin{figure*}
    \centering
    \includegraphics{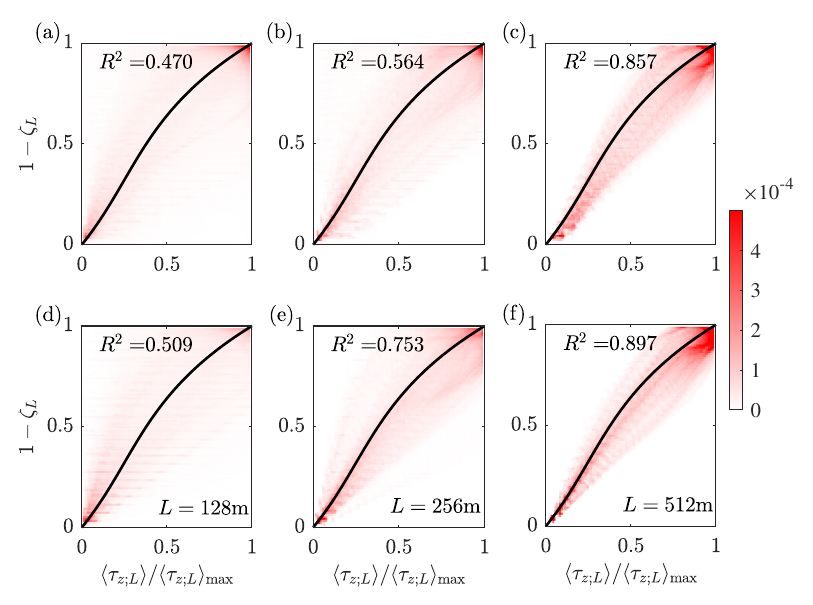}
    \caption{The distribution density of normalised local cumulative turbulent stress $\sav{\tau_{z; L}}/\sav{\tau_{z; L}}_{\max}$ against the local scaled frontal area $\zeta_L$ at different averaging lengths. (a, d) $L =128$ m, (b, e) $L=256$ m, (c, f) $L= 512$ m, for CC (top panel) and SC (bottom panel), respectively. The solid line represents the parametrisation Eq. \eqref{eq:tau_z_3d} with modified constants $A = 0.89, \, B = 1.82$. The coefficient of determination $R^2$ is labelled to show the performance of the parametrisation.}
    \label{Fig9}
\end{figure*}

Figure \ref{Fig9} presents the performance of the total kinematic stress parameterisation (Eq. \eqref{eq:tau_z_3d}) with the modified constants. Although the parameterisation reproduces the plane-averaged profiles reasonably well (see Figs. \ref{Fig3}i,j), its accuracy declines as the resolution increases. Only at the coarsest resolution ($L = 512$ m) does the parameterisation provide an acceptable fit for both cases.

The primary source of deviation lies in the constant-force assumption applied within the canopy, namely $\tau_{f;L}(z<h_{\max}) \approx \tau_{0;L} \approx \sav{\tau_{z;L}}_{\max}$. This assumption is adequate in the plane-averaged sense, i.e., $L = \infty$, where variations in the cumulative forcing are negligible relative to its magnitude. However, when applied locally within a filter, the forcing magnitude is reduced, making the relative contribution of its variation more significant and thus undermining the assumption. Furthermore, combining the horizontal transport effect, the deviation is further magnified at higher resolutions.

\section{Conclusions} \label{sec: conclusions}

A central question for next-generation urban canopy modelling is whether the standard assumption of horizontal homogeneity remains defensible as NWP grids approach hectometric resolution \citep{Barlow2014, Lean2024}. At these scales, urban surfaces enter the so-called \emph{building grey zone}, in which large buildings and the flow around them become partially resolved while smaller-scale variability does not, so that surface parametrisations become more complex rather than simpler and horizontal exchanges between adjacent cells become first-order \citep{Lean2024}. This work addresses the question directly by providing a diagnostic, the characteristic urban length scale $\ell$, that pinpoints, for any given morphology, the resolution at which resolved and unresolved heterogeneity become equally important. $\ell$ separates two distinct modelling regimes: at $L \gtrsim \ell$, the flow is sufficiently homogenised that conventional plane-averaged closures remain useful; at $L < \ell$, horizontal transport and filter-to-filter variability are first-order, and scale-aware (or heterogeneity-aware) formulations become necessary.

We obtained $\ell$ by applying the framework of \citet{MVR2025}, a 2-D horizontal coarse-graining filter that generates a series of coarser fields and quantifies the resolved and unresolved variances of the flow at each scale, to building-resolved LES of two layouts of the University of Bristol campus, the same site used in the ASSURE wind-tunnel programme \citep{Bi2025}. In the disc configuration (CC), the buildings form an approximately circular cluster with large open spaces at the corners of the periodic patch; in the modified layout (SC), these corners are infilled with additional buildings. The diagnosed length scale is strongly morphology-dependent: $\ell \approx 256$ m for CC, about one third of the domain size, which is set by the radius of the cluster and reflecting a neighbourhood-scale heterogeneity consistent with the few-hundred-metre variability scale typically associated with urban morphology \citep{Lean2024}. This scale is one that current kilometre-scale NWP cannot resolve and that even hectometric models would only partially capture. For SC, $\ell \approx 64$ m, the scale of individual buildings. Although based on two related morphologies, the four-fold contrast already establishes that $\ell$ is not a universal property of cities but a layout-specific diagnostic that should be computed before any horizontally averaged closure is applied.

An \emph{a priori} test of the distributed drag and turbulent-stress parametrisations of \citet{MVR2025}, with constants refit to the present cases ($A = 0.89$, $B = 1.82$), confirms the role of $\ell$ as a transition scale. These parametrisations, derived from idealised morphologies under a horizontally homogeneous assumption, perform well at NWP-like resolutions $L \gtrsim \ell$ but lose fidelity rapidly at finer scales, where the underlying assumption is violated and horizontal exchanges between cells can no longer be neglected. The drag parametrisation degrades more gracefully; the turbulent-stress parametrisation, which additionally invokes a constant-stress canopy assumption, fails earlier. The decline is steeper than that observed in idealised cuboid arrays, reflecting the much larger filter-to-filter variability of realistic layouts and helping to explain why parametrisations transferred unchanged from idealised studies often disappoint when applied to real cities.

The principal implication is that the resolution dependence of urban closures cannot be addressed without first asking what scale of heterogeneity the morphology actually contains. The framework employed offers a practical \emph{a priori} tool to assess the scale awareness of urban parametrisations from morphology alone, and provides a quantitative basis on which scale-aware closures, whose form and coefficients adapt to $L/\ell$, can be developed, directly addressing the call by \citet{Lean2024} for LES-based development of urban parametrisations on realistic geometries. The remaining steps are to apply the diagnostic across a wider range of cities and local-climate-zone types, to extend it beyond neutral stability, and to translate it into operational closures that bridge kilometre-scale and hectometric NWP.

\section*{Acknowledgements}
The support of the ARCHER2 UK National Supercomputing Service (project ARCHER2-eCSE05-3) and the NERC highlight grant ASSURE: Across-Scale ProcesseS in Urban Environment (NE/W002868/1) is acknowledged. 

\section*{Data availability}
The data that support the findings of this study are available from the corresponding author upon reasonable request.

\section*{Declaration of interests}
The authors report no conflict of interest.

\appendix

\section{Computational details}
The convolution in Eq. \eqref{eq:convolution} can be efficiently computed using two-dimensional Fast Fourier Transforms (2-D FFTs). For ease of implementation, the equation can be rewritten by introducing a fluid mask $I_f$ ($1$ in the fluid region and $0$ in the solid), which extends the integral over the fluid region to the entire plane $\Omega(z)$.
\begin{equation}
 \sav{\varphi}_L(\vec x) =  \int_{\Omega(z)} \mathcal A(\vec x_\perp - \vec y_\perp)  I_f(\vec y_\perp, z) \varphi(\vec y_\perp, z) \, \d \vec y_\perp \, ,
\end{equation}
Specifically, the horizontal FFTs of $\varphi$ and the kernel $\mathcal A$ are first computed. Their product is then transformed back to physical space via the inverse FFT, yielding the filtered field at all horizontal points simultaneously. This approach reduces computational cost from direct spatial convolution to approximately $O((\log n)^2 n^3)$ operations for a cubic domain of size $n^3$, which is particularly beneficial to large domains and large filter lengths. The computational details can be found at \citet{MVR2025Zenodo}.

A key feature of uDALES is its ability to represent solid surfaces through a facet mesh through the immersed boundary method. Each facet, denoted $m$, is associated with a Cartesian grid cell at indices ($I_m, J_m, K_m$). A surface quantity $\phi_m$ on facet $m$ with area $A_m$ can thus be mapped to a volumetric density field $\rho_\phi$, which is non-zero only in the cells adjacent to the surface boundary. In discrete form,
\begin{equation}
  \label{eq:Phi}
  \rho_{\phi;ijk} = \sum_{m \in M_{ijk}} \frac{\phi_m A_m}{\Delta x \Delta y \Delta z_{K_m}} \, ,
\end{equation}
where $M_{ijk}$ is the set of all facets associated with Cartesian cell $(i, j, k)$. This mapping allows surface fluxes to be consistently incorporated into the LES framework.

Integrating over the solid–fluid interface then provides the total flux exchanged across the boundary, any surface quantity can be filtered straightforwardly using
\begin{equation}
    \sav{\rho_\phi}(\vec x) = \int_{\Omega_f} \mathcal A(\vec x_\perp - \vec y_\perp) \rho_\phi(\vec y_\perp) \d \vec y_\perp
\end{equation}
The distributed drag term $\sav{f_D}$ can be evaluated by choosing $\phi_m = -(\overline p_m \vec e_x  - \nu (\nabla \overline u)_m) \cdot \vec N_m$, which includes the surface pressure (form drag) and viscous shear stress (skin drag) contributions in the streamwise direction. Denoting the volumtric drag density by $\sav{\rho_D}$, we have $\sav{f_D} = \sav{\rho_D}$.

The frontal area density can be evaluated by setting $\phi_m = - \min(\vec e_u \cdot \vec N_m, 0)$, where $e_u$ is the unit vector of the local velocity. This yields a frontal-area density field $\rho_L$ such that
$\int \rho_L \d x \d y = L_b$ and $\int \rho_L \d V = A_F$.
Recall that $L_b$ is the frontal building width. From this, the filter-dependent frontal area index is defined as
\begin{equation} 
\zeta_L(\vec x_\perp, z) = \frac{1}{\lambda_{f;L}} \int_z^h \sav {\rho_{L}}_L(\vec x_\perp, z') \, \d z' \, , \end{equation}
with
\begin{equation} 
\lambda_{f;L}(\vec x_\perp) = \int_0^h \sav {\rho_{L}}_L(\vec x_\perp, z') \, \d z' \, . 
\end{equation}
\bibliographystyle{wileyqj.bst}
\bibliography{References.bib}

\end{document}